\pdfoutput=1

\documentclass[11pt]{article}

\usepackage[final]{acl}

\usepackage{times}
\usepackage{latexsym}

\usepackage[T1]{fontenc}

\usepackage[utf8]{inputenc}

\usepackage{microtype}

\usepackage{inconsolata}

\usepackage{algorithm}
\usepackage{algorithmic}
\usepackage{multirow}
\usepackage{booktabs}
\usepackage{amsmath}
\usepackage{amssymb}
\usepackage{graphicx}

\usepackage{newfloat}
\usepackage{listings}

\title{TCSinger 2: Customizable Multilingual Zero-shot Singing Voice Synthesis}

\author{%
\normalsize
Yu Zhang\thanks{Equal contribution}\quad
Wenxiang Guo\footnotemark[1]\quad
Changhao Pan\footnotemark[1]\quad
Dongyu Yao\quad
\\\normalsize
\textbf{Zhiyuan Zhu}\quad
\textbf{Ziyue Jiang}\quad
\textbf{Yuhan Wang}\quad
\textbf{Tao Jin}\quad
\textbf{Zhou Zhao}\thanks{Corresponding Author}\quad
\\
Zhejiang University\\
\texttt{\{yuzhang34,zhaozhou\}@zju.edu.cn}
}

\begin{document}
\maketitle
\begin{abstract}

Customizable multilingual zero-shot singing voice synthesis (SVS) has various potential applications in music composition and short video dubbing. 
However, existing SVS models overly depend on phoneme and note boundary annotations, limiting their robustness in zero-shot scenarios and producing poor transitions between phonemes and notes. 
Moreover, they also lack effective multi-level style control via diverse prompts. 
To overcome these challenges, we introduce TCSinger 2, a multi-task multilingual zero-shot SVS model with style transfer and style control based on various prompts. 
TCSinger 2 mainly includes three key modules: 
1) Blurred Boundary Content (BBC) Encoder, predicts duration, extends content embedding, and applies masking to the boundaries to enable smooth transitions.
2) Custom Audio Encoder, uses contrastive learning to extract aligned representations from singing, speech, and textual prompts.
3) Flow-based Custom Transformer, leverages Cus-MOE, with F0 supervision, enhancing both the synthesis quality and style modeling of the generated singing voice.
Experimental results show that TCSinger 2 outperforms baseline models in both subjective and objective metrics across multiple related tasks. 
Singing voice samples are available at \url{https://aaronz345.github.io/TCSinger2Demo/}.
Code can be found at \url{https://github.com/AaronZ345/TCSinger2}.

\end{abstract}

\section{Introduction}

Zero-shot singing voice synthesis (SVS) aims to generate high-quality singing voices with unseen multi-level styles based on audio or textual prompts \citep{zhang2024stylesinger,zhang2024tcsinger,guo2025techsinger}. 
This field has found widespread potential applications in professional music composition and short video dubbing. 
Zero-shot SVS involves using an acoustic model to leverage lyrics and musical notations for content modeling, while audio or textual prompts can control singing styles. 
Finally, a vocoder is employed to synthesize the target singing voice.

Although traditional SVS tasks \citep{zhang2022visinger, kim2022adversarial, cho2022mandarin} have made significant strides, there is an increasing demand for more \textbf{customizable} experiences. This includes not only \textbf{zero-shot style transfer} by audio prompts \citep{du2024cosyvoice}, but also the need to leverage natural language textual prompts for \textbf{multi-level style control}. 
Textual prompts can influence global timbre by specifying the singer's gender and vocal range. 
Additionally, they can control broader aspects of singing style, such as vocal techniques (e.g., bel canto) and emotional expression (e.g., happy or sad), as well as segment- or word-level techniques (e.g., mixed voice or falsetto). 
Audio prompts, in addition, enable the target to learn these consistent multi-level styles while incorporating accent, pronunciation, and transitions. 
However, current models still struggle to effectively implement style transfer and style control based on various prompts in zero-shot scenarios. Consequently, achieving a natural, stable, and highly controllable generation remains a significant challenge.

Currently, customizable multilingual zero-shot SVS faces two major challenges:
1) Existing SVS models heavily rely on phoneme and note boundary annotations, which limits their robustness. 
Datasets like OpenCpop \citep{wang2022opencpop} depend on MFA and human-ear alignment, which introduces significant errors at the boundaries. 
Additionally, these SVS models often produce poor transitions between phonemes and notes, especially in zero-shot scenarios, where this issue becomes even more pronounced. 
\citet{choi2022melody} introduces a melody-unsupervised model to reduce reliance on boundary annotations. 
However, the unsupervised approach results in lower synthesis quality and cannot ensure smooth transitions at the boundaries.
2) Existing SVS models with style transfer and style control lack effective multi-level style control through diverse prompts. 
TCSinger \citep{zhang2024tcsinger} achieves style control using specified labels or audio prompts. 
However, it still cannot cover a wider range of applications with more flexible prompts, including natural language textual, speech, or singing prompts. 
Moreover, its capability in style control is still quite limited.

To address these challenges, we introduce TCSinger 2, a multi-task multilingual zero-shot SVS model with style transfer and style control based on various prompts.
TCSinger 2 enables effective style control using natural language textual, speech, or singing prompts.
To achieve smooth and robust phoneme/note boundary modeling, we design the Blurred Boundary Content (BBC) Encoder. 
This encoder predicts duration, extends content embedding, and applies masking to phoneme and note boundaries to facilitate smooth transitions and ensure robustness.
Furthermore, to extract aligned representations from singing, speech, and textual prompts, we propose the Custom Audio Encoder based on contrastive learning, extending the model's applicability to a broader range of related tasks.
In addition, to generate high-quality and highly controllable singing voices, we introduce the Flow-based Custom Transformer. 
Within this framework, we utilize Cus-MOE, which, depending on the language and textual or audio prompt, selects different experts to achieve better synthesis quality and style modeling.
Moreover, we incorporate additional supervision using F0 information to enhance the expressiveness of the synthesized output.
Our experimental results show that TCSinger 2 outperforms other baseline models in synthesis quality, singer similarity, and style controllability across various tasks, including zero-shot style transfer, cross-lingual style transfer, multi-level style control, and speech-to-singing (STS) style transfer.
Our main contributions can be summarized as:

\begin{itemize}
\item We present TCSinger 2, a multi-task multilingual zero-shot SVS model with style transfer and style control based on various prompts. 
\item We introduce the Blurred Boundary Content Encoder for robust modeling and smooth transitions of phoneme and note boundaries.
\item We design the Custom Audio Encoder using contrastive learning to extract styles from various prompts, while the Flow-based Custom Transformer with Cus-MOE and F0, enhances synthesis quality and style modeling.
\item Experimental results show that TCSinger 2 outperforms baseline models in subjective and objective metrics across multiple tasks.
\end{itemize}

\section{Related Works}
\paragraph{Singing Voice Synthesis.}
Singing Voice Synthesis (SVS) focuses on generating high-quality singing voices from lyrics and musical notes. 
VISinger 2 \citep{zhang2022visinger} enhances synthesis quality by employing digital signal processing techniques. 
SiFiSinger \citep{cui2024sifisinger} extends VISinger by improving pitch control with a source module that generates F0-controlled excitation signals. 
Additionally, MuSE-SVS \citep{kim2023muse} introduces a multi-singer emotional singing voice synthesizer, enhancing expressiveness. 
For singing datasets, Opencpop \citep{wang2022opencpop} and GTSinger \citep{zhang2024gtsinger} have made significant contributions by releasing annotated datasets. 
More recently, TCSinger \citep{zhang2024tcsinger} introduces an adaptive normalization method that enhances the details in synthesized voices. 
Despite the strong performance of these models in generation quality, they typically require precise alignment of audio, lyrics, and notes, which is limited by the quality of the dataset itself and leads to unnatural transitions at phoneme and pitch boundaries, particularly evident in zero-shot scenarios. 
To address this, we employ blurred boundaries.

\begin{figure*}[t]
\centering
\includegraphics[width=1.0\textwidth]{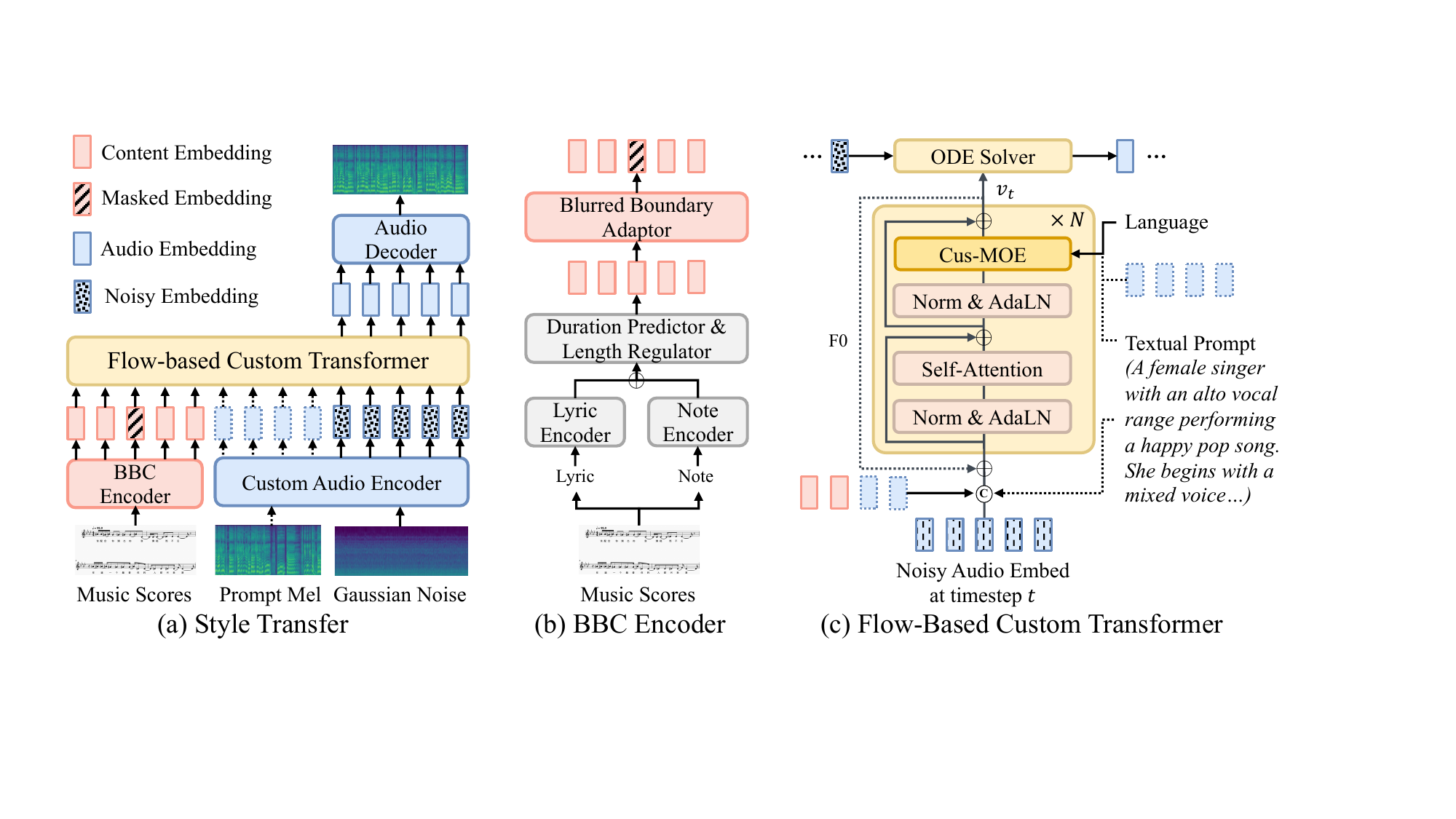}
\caption{
The architecture of TCSinger 2.
BBC Encoder denotes Blurred Boundary Content Encoder.
Figure (a) shows the style transfer process.
Either mel from audio prompt or textual prompt can control multi-level styles.
}
\label{fig: arch}
\end{figure*}

\paragraph{Style Modeling.}
Style modeling is crucial for generating expressive singing voices in a controlled manner, typically involving the transfer of styles from reference audio \citep{wagner2010experimental}. 
\citet{skerry2018towards} is the first to integrate a style reference encoder into a Tacotron-based TTS system, enabling the transfer of style for similar-text speech. 
Attentron \citep{choi2020attentron} introduces an attention mechanism to extract styles from reference samples. 
ZSM-SS \citep{kumar2021normalization} proposes a Transformer-based architecture with an external speaker encoder using wav2vec 2.0 \citep{baevski2020wav2vec}. 
Daft-Exprt \citep{zaidi2021daft} employs a gradient reversal layer to improve target speaker fidelity in style transfer. 
StyleTTS 2 \citep{li2024styletts} predicts pitch and energy based on a prosody predictor \citep{li2022styletts}, while CosyVoice \citep{du2024cosyvoice} incorporates x-vectors into an LLM to model and disentangle styles. 
PromptSinger \citep{wang2024prompt} attempts to control speaker identity based on text descriptions. 
Although these methods can model certain aspects of styles, they are unable to model multi-level singing styles using natural language textual prompts, as well as achieve greater customizability by multilingual speech and singing prompts.

\section{Method}

In this section, we first provide an overview of the proposed TCSinger 2.
Next, we describe several key components, including the Blurred Boundary Content (BBC) Encoder, Custom Audio Encoder, and Flow-based Custom Transformer. 
Finally, we detail the training and inference processes.

\subsection{Overview}

The architecture of TCSinger 2 is shown in Figure \ref{fig: arch}(a). 
Let $y_{gt}$ represent the ground truth singing voice, and $m_{gt} \in \mathbb{R}^{80 \times T}$ represent the mel spectrogram, where $T$ denotes the target length. 
The Custom Audio Encoder compresses $m_{gt}$ into $\hat{m_{gt}}$, and the generation process is given by $G(\epsilon \mid C,P) \xrightarrow{} \hat{m_{pr}} \xrightarrow{} m_{gt}$, where $\epsilon$ is Gaussian noise and $C$ represents the conditions.
$C$ includes the lyrics $l$ and music notation $n$ extracted from the music scores. 
$P$ can be one of singing prompt $p_{si}$, speech prompt $p_{sp}$, and textual prompt $p_{te}$. 
The lyrics $l$ and notation $n$ are inputted to the BBC Encoder, which predicts the duration, and extends the content embedding. 
It also applies masking at the boundaries to facilitate smooth transitions and ensure robustness, producing $z_c$.
The Custom Audio Encoder utilizes contrastive learning to extract consistent representations from singing, speech, and textual prompts. 
When transferring styles from audio prompt $p_a$ ($p_{si}$ or $p_{sp}$), it extracts a style-rich representation $ z_{pa}$. 
When using textual prompt $p_{te}$ for style control, it is encoded into multi-style controlling representation $z_{pt}$.
Finally, the Flow-Based Custom Transformer leverages $z_c$, $z_t$, as well as $z_{pt}$ or $z_{pa}$ to generate the predicted singing voice $y_{pr}$.

\begin{figure*}[t]
\centering
\includegraphics[width=1.0\textwidth]{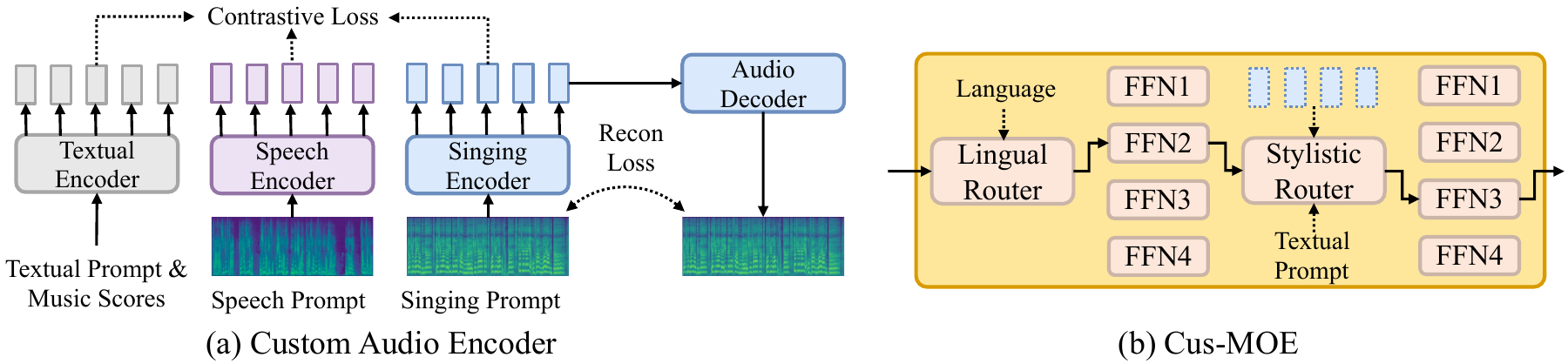}
\caption{
The architecture of Custom Audio Encoder and Cus-MOE.
In Figure (a), different encoders extract aligned representations based on the input. 
In Figure (b), each router selects one FFN based on conditions during inference.
}
\label{fig: arch2}
\end{figure*}

\subsection{BBC Encoder}

Current SVS models rely heavily on precise phoneme and note boundary annotations, which are often automated using tools like MFA. 
However, manual post-editing datasets are rare, and even those based on human auditory annotations contain many errors \citep{wang2022opencpop, zhang2024gtsinger}. 
This is particularly problematic in multilingual singing datasets, where annotation errors and data scarcity lead to mislearning of phonemes and pitch. 
For example, when the latter half of a phoneme’s duration actually belongs to the next phoneme, the model struggles to learn the pronunciation of both phonemes correctly. 
Additionally, current SVS models produce poor transitions between phonemes and notes, particularly in zero-shot scenarios, where this issue is more pronounced.

To address this issue and simultaneously expand the dataset while enhancing the naturalness and musicality of transitions in zero-shot settings, we introduce the Blurred Boundary Content (BBC) Encoder.
As shown in Figure \ref{fig: arch} (b), after separately encoding the lyrics $l$ and notes $n$, we predict the duration and extend the content embedding, resulting in a frame-level sequence $[z_{c1}, z_{c1}, z_{c2}, z_{c2}, \dots, z_{cn}]$ with precise boundaries.
Next, we randomly mask $m$ tokens at each phoneme and note boundaries to produce $[z_{c1}, \varnothing, z_{c2}, z_{c2}, \varnothing, \dots, z_{cn}]$. 
By adjusting $m$, we can strike a balance between providing more supervision and achieving better robustness.
Considering our compression rate and sample rate, we set $m = 8$. 
Note that $m$ will not cover too short contents.
With the BBC Encoder, we obtain blurred boundaries, then refined in the Flow-based Custom Transformer, where self-attention mechanisms establish fine-grained implicit alignment paths.
The BBC Encoder expands the roughly aligned dataset, improves the naturalness of transitions, and enhances the quality of zero-shot generation.

\subsection{Custom Audio Encoder}

The style of singing is very complex, encompassing factors such as timbre, singing method, emotion, technique, accent, and more. 
This makes it challenging to compress the singing voice mel while extracting a representation that is rich in multi-level style. 
Such a representation is crucial for both style transfer and style control. Additionally, to expand the customizable application scenarios, it is important to extract an aligned style representation from speech as well. 
This allows users to produce singing voices that match their speech style.

As shown in Figure \ref{fig: arch2} (a), based on the singing prompt $p_{si}$, speech prompt $p_{sp}$, and textual prompt $p_{te}$ with content $C$, we extract a triplet pair $(z_{psi}, z_{psp}, z_{ptc})$. 
We also conduct reconstruction to ensure $z_{psi}$ does not compromise the integrity of singing voices. 
The singing and speech encoders, and the audio decoder, are based on the VAE model \citep{kingma2013auto}. 
For the textual encoder, we use cross-attention to combine music scores and textual prompts, obtaining a representation with content and multi-level styles. 
We use contrastive learning to align the triplet pairs, ensuring they all contain unified styles.
We design three types of contrasts: (1) same content, different styles; (2) similar styles, different content; and (3) different styles and contents. 
We use the contrastive objective \citep{radford2021learning} for training:
\begin{equation}
\begin{aligned}
\label{equ: contras}
& \mathcal{L}_{p_{si}^i,p_{sp}^i} = \log \frac{\exp ({sim}({z_{si}}^i, {z_{sp}}^i)/\tau)}{\sum_{j=1}^{N} \exp ({sim}({z_{si}}^i, {z_{sp}}^j)/\tau)} \\
& + \log \frac{\exp ({sim}({z_{sp}}^i, {z_{si}}^i)/\tau)}{\sum_{j=1}^{N} \exp ({sim}({z_{sp}}^i, {z_{si}}^j)/\tau)},
\end{aligned}
\end{equation}
where $\text{sim}(\cdot)$ denotes cosine similarity. 
The total loss $\mathcal{L}_{\text{contras}} = -\frac{1}{6N}\sum_{i=1}^{N} (\mathcal{L}_{p_{si}, p_{sp}} + \mathcal{L}_{p_{sp}^i, p_{te}^i} + \mathcal{L}_{p_{si}^i, p_{te}^i})$. 
Therefore, three embeddings are aligned in the same space.
To train the Audio Decoder, we use L2 loss $\mathcal{L}_{\text{recon}}$ and LSGAN-style adversarial loss $\mathcal{L}_{\text{adv}}$ \citep{mao2017least} with a GAN discriminator for better reconstruction.
The textual encoder supervises styles and contents, enriching the audio embedding with styles without losing content.
For more details, please refer to Appendix \ref{sec: appendix1vae}.

\subsection{Flow-based Custom Transformer}

\paragraph{Flow-based Transformer.}
Singing voices are highly complex and stylistically diverse, making modeling particularly challenging. To address this, we propose the Flow-based Custom Transformer. As shown in Figure \ref{fig: arch} (c), we combine the flow-matching technique, which can generate stable and smooth paths, to achieve robust and fast inference. Additionally, we leverage the sequence learning ability of the transformer's attention mechanism to improve the quality and style modeling of SVS.

During training, we add Gaussian noise $\epsilon$ to the audio encoder's output $\hat{m_{gt}}$ to obtain $x_t$ at timestep $t$, which is achieved via linear interpolation. We then concatenate $x_t$ with the content embedding $z_c$ from the BBC Encoder, and an optional audio prompt embedding $z_{pa}$ (either $z_{psi}$ or $z_{psp}$) from the Custom Audio Encoder. 
This allows the model to use self-attention to learn content and style transfer.
When using natural language textual prompts to control styles, we also encode it as $z_{pt}$ and concatenate instead of $z_{pa}$ to achieve multi-level style control.
Furthermore, we employ RMSNorm \citep{zhang2019root} and AdaLN \citep{peebles2023scalable} to ensure training stability and global modulation with styles and timestep. 
RoPE \citep{su2024roformer} is also used to enhance the model's ability to capture dependencies across sequential frames.
The output vector field of our model in each $t$ is trained with the flow-matching objective:
\begin{equation}
\label{equ: loss}
\begin{aligned}
&\mathcal{L}_{flow} = \mathbb{E}_{t, p_t(x_{t})} \left\| v_t(x_t, t | C; \theta) - (\hat{m_{gt}} - \epsilon) \right\|^2,
\end{aligned}
\end{equation}
where $p_t(x_{t})$ represent the distribution of $x_{t}$ at timestep $t$.
Additionally, given the importance of pitch in singing styles \citep{zhang2024stylesinger}, we use the first block's output to predict F0, providing supervision and input for subsequent blocks. 
During inference, $\epsilon$ is combined with the condition to generate the target $\hat{m_{pr}}$ with fewer timesteps than in training, resulting in a smoother generation.
For more details, please refer to Appendix \ref{sec: appendix1flow}.

\paragraph{Cus-MOE.}

To achieve higher-quality multilingual generation and better style modeling, we propose Cus-MOE (Mixture of Experts), selecting suitable experts based on various conditions. 
As shown in Figure \ref{fig: arch2} (b), our Cus-MOE consists of two expert groups, each focusing on linguistic and stylistic conditions.
The Lingual-MOE selects experts based on lyric languages, with each expert specializing in a particular language family (such as Latin), using domain-specific experts to improve generation quality for each language family.
The Stylistic-MOE conditions on audio or natural language textual prompts, adjusting inputs to match fine-grained styles, such as an expert specializing in alto range female and happy pop falsetto singing.

Our routing strategies use a dense-to-sparse Gumbel-Softmax \citep{nie2021evomoe}, which reparameterizes categorical variables to make sampling differentiable, enabling dynamic routing.
Let $h$ be the hidden representation, and $g(h)_i$ denote the routing score for expert $i$. 
To prevent overloading, we apply a load-balancing loss \citep{fedus2022switch}:
\begin{equation}
\label{equ: balance}
\begin{aligned}
&\mathcal{L}_{balance} = \alpha N \sum_{i=1}^{N} \left( \frac{1}{B} \sum_{h \in B} g(h)_i \right),
\end{aligned}
\end{equation}
where $B$ is the batch size, $N$ is the number of experts, and $\alpha$ controls regularization strength.
For more details, please refer to Appendix \ref{sec: appendix1moe}.

\subsection{Training and Inference Procedures}

\paragraph{Training Procedures} 
For the pre-trained custom audio encoder and decoder, the final loss includes:
1) $\mathcal{L}_{contras}$: the contrastive objective for contrastive learning;
2) $\mathcal{L}_{rec}$: the L2 reconstruction loss;
3) $\mathcal{L}_{adv}$: the LSGAN-styled adversarial loss in GAN discriminator.
For TCSinger 2, the final loss terms during training consist of the following aspects:
1) $\mathcal{L}_{dur}$: the mean squared error (MSE) phoneme-level duration loss on a logarithmic scale in the BBC Encoder;
2) $\mathcal{L}_{pitch}$: the MSE pitch loss in the log scale.
3) $\mathcal{L}_{balance}$: the load-balancing loss for each expert group in Cus-MOE;
4) $\mathcal{L}_{flow}$: the flow matching loss of Flow-based Custom Transformer.

\paragraph{Inference Procedures}
TCSinger 2 supports multiple inference tasks based on the input prompt. For unseen singing prompts, it performs zero-shot style transfer, whether the content and prompt are in the same language or across languages. If the input includes lyrics and a singing prompt in different languages, the model can perform cross-lingual style transfer. Given a natural language textual prompt, TCSinger 2 enables multi-level style control. 
When provided with a speech prompt, it can carry out speech-to-singing (STS) style transfer.

To enhance generation quality and style controllability, we incorporate the classifier-free guidance (CFG) strategy. 
During training, we randomly drop input prompts with a probability of 0.2. 
During inference, we modify the output vector field as:
\begin{equation}
\label{equ: cfg}
\begin{aligned}
& v_{cfg}(x, t|C,P;\theta) = \gamma v_t(x, t|C,P;\theta) +\\
& (1-\gamma) v_t(x,|C,\varnothing;\theta),
\end{aligned}
\end{equation}
where $\gamma$ is the CFG scale that balances creativity and controllability. 
We set $\gamma = 3$ to improve generation quality and enhance style control. 
Finally, by leveraging the accelerated inference capabilities of the flow-matching method, our model can efficiently and robustly generate singing voices.

\section{Experiments}

\subsection{Experimental Setup}

\begin{table*}[t]
\centering
\small
\scalebox{1}{
\begin{tabular}{lcccccc}
\toprule
\multirow{2}{*}{\bfseries{Method}} & \multicolumn{4}{c}{\bfseries{Parallel}} & \multicolumn{2}{c}{\bfseries{Cross-Lingual}}\\ \cmidrule(lr){2-5} \cmidrule(lr){6-7}
 & MOS-Q $\uparrow$ & MOS-S $\uparrow$ & FFE $\downarrow$ & Cos $\uparrow$ & MOS-Q $\uparrow$ & MOS-S $\uparrow$ \\
\midrule  
GT & 4.58 $\pm$ 0.11 & / & / & / & / & /
\\
GT (vocoder) & 4.36 $\pm$ 0.08 & 4.41 $\pm$ 0.13 & 0.04 & 0.95 & / & / \\
\midrule  
StyleTTS 2 & 3.71 $\pm$ 0.14 & 3.79 $\pm$ 0.09 & 0.42 & 0.71 & 3.58 $\pm$ 0.16 & 3.63 $\pm$ 0.12\\
CosyVoice & 3.74 $\pm$ 0.10 & 3.93 $\pm$ 0.15 & 0.33 & 0.87 & 3.63 $\pm$ 0.08 & 3.77 $\pm$ 0.17\\
VISinger 2 & 3.79 $\pm$ 0.17 & 3.88 $\pm$ 0.11 & 0.31 & 0.83 & 3.69 $\pm$ 0.19 & 3.72 $\pm$ 0.06\\
TCSinger& 3.94 $\pm$ 0.06 & 4.01 $\pm$ 0.18 & 0.26 & 0.91 & 3.77 $\pm$ 0.13 & 3.87 $\pm$ 0.14\\
\midrule  
TCSinger 2 (ours) & \bf 4.13 $\pm$ 0.12 & \bf 4.27 $\pm$ 0.09 & \bf 0.21 & \bf 0.93 & \bf 3.96 $\pm$ 0.10 & \bf 4.09 $\pm$ 0.07\\
\bottomrule      
\end{tabular}}
\caption{
Synthesis quality and singer similarity of zero-shot parallel and cross-lingual style transfer. 
}
\label{tab: base}
\end{table*}

\paragraph{Dataset.}

The dataset for singing voices is quite limited. 
However, using the blurred boundary strategy, we expand our dataset by collecting 50 hours of clean singing voices and annotating them.
Then, we use several open-source singing datasets, including Opencpop \citep{wang2022opencpop} (Chinese, 1 singer, 5 hours of singing voices), M4Singer \citep{zhang2022m4singer} (Chinese, 20 singers, 30 hours of singing voices), OpenSinger \citep{huang2021multi} (Chinese, 93 singers, 85 hours of singing voices), PopBuTFy \citep{liu2022learning} (English, 20 singers, 18 hours of speech and singing voices), and GTSinger \citep{zhang2024gtsinger} (9 languages, 20 singers, 80 hours of singing and speech).
All languages include Chinese, English, French, Spanish, German, Italian, Japanese, Korean, and Russian.
We manually annotate part of these data with multi-level style labels (like emotions).
Then, we randomly select 30 singers as the unseen test set to evaluate zero-shot performance for all tasks. 
Our dataset partitioning carefully ensures that training and test sets for all tasks contain multilingual speech and singing data.
For more details, please refer to Appendix \ref{sec: appendix2}.

\paragraph{Implementation Details.}
We set the sample rate to 48,000 Hz, the window size to 1024, the hop size to 256, and the number of mel bins to 80 to derive mel-spectrograms from raw waveforms. 
The output mel-spectrograms are transformed into singing voices by a pre-trained HiFi-GAN vocoder \citep{kong2020hifi}.
We utilize four Transformer blocks as the vector field estimator. 
Each Transformer layer employs a hidden size of 768 and eight attention heads. 
The Cus-MoE includes four experts per expert group. 
During training, flow-matching uses 1,000 timesteps, while inference uses 25 timesteps with the Euler ODE solver.
We train all our models with eight NVIDIA RTX-4090 GPUs. 
For more model details, please refer to Appendix \ref{sec: appendix1arch}.

\paragraph{Evaluation Details.}
We use both objective and subjective evaluation metrics to validate the performance of TCSinger 2. 
For subjective metrics, we conduct the MOS (mean opinion score) evaluation.
We employ the MOS-Q to judge synthesis quality (including fidelity, clarity, and naturalness), MOS-S to assess singer similarity (in timbre and other styles) between the result and prompt, and MOS-C to evaluate controllability (accuracy and expressiveness of style control).
Both these metrics are rated from 1 to 5 and reported with 95\% confidence intervals. 
For objective metrics, we use Singer Cosine Similarity (Cos) to judge singer similarity, and F0 Frame Error (FFE) to quantify synthesis quality. 
For more details, please refer to Appendix \ref{sec: appendix3}.

\paragraph{Baseline Models.}

We conduct a comprehensive comparative analysis of synthesis quality, style controllability, and singer similarity for TCSinger 2 against several baseline models. Initially, we evaluate our model against the ground truth (GT) and the audio generated by the pre-trained HiFi-GAN (GT (vocoder)).
We first compare it with two strong zero-shot multilingual speech synthesis baseline models, including StyleTTS 2 \citep{li2024styletts} and CosyVoice \citep{du2024cosyvoice}. To ensure a fair comparison for singing tasks, we enhance these models with a note encoder to process musical notations and train them on our multilingual speech and singing data.
Next, we select a traditional high-fidelity SVS model, VISinger 2 \citep{zhang2022visinger}, and the first zero-shot SVS model with style transfer and style control, TCSinger \citep{zhang2024tcsinger}. 
We employ their open-source codes.
For more details, please refer to Appendix \ref{sec: appendix4}.

\subsection{Main Results}

\begin{table*}[t]
\centering
\small
\scalebox{1}{
\begin{tabular}{lccccc}
\toprule
\multirow{2}{*}{\bfseries{Method}} & \multicolumn{3}{c}{\bfseries{Parallel}} & \multicolumn{2}{c}{\bfseries{Non-Parallel}}\\ \cmidrule(lr){2-4} \cmidrule(lr){5-6}
& {MOS-Q $\uparrow$} & {MOS-C $\uparrow$} & {FFE $\downarrow$} & {MOS-Q $\uparrow$} & {MOS-C $\uparrow$} \\
\midrule
GT & 4.56 $\pm$ 0.13 & / & / & / & / \\
GT (vocoder) & 4.26 $\pm$ 0.09 & 4.32 $\pm$ 0.11 & 0.06 & / & / \\
\midrule
StyleTTS 2 & 3.61 $\pm$ 0.18 & 3.67 $\pm$ 0.14 & 0.43 & 3.51 $\pm$ 0.16 & 3.59 $\pm$ 0.07 \\
CosyVoice & 3.72 $\pm$ 0.07 & 3.73 $\pm$ 0.10 & 0.37 & 3.60 $\pm$ 0.19 & 3.67 $\pm$ 0.13\\
VISinger 2 & 3.81 $\pm$ 0.15 & 3.81 $\pm$ 0.06 & 0.30 & 3.69 $\pm$ 0.08 & 3.75 $\pm$ 0.12 \\
TCSinger & 3.99 $\pm$ 0.12 & 3.97 $\pm$ 0.08 & 0.27 & 3.90 $\pm$ 0.14 & 3.93 $\pm$ 0.10 \\
\midrule
TCSinger 2 (ours) & \bf 4.07 $\pm$ 0.10 & \bf 4.19 $\pm$ 0.16 & \bf 0.22 & \bf 3.98 $\pm$ 0.11 & \bf 4.11 $\pm$ 0.09 \\
\bottomrule
\end{tabular}}
\caption{
Multi-level style control performance in parallel and non-parallel experiments based on textual prompts.
}
\label{tab: sc}
\end{table*}

\begin{figure*}[t]
\centering
\includegraphics[width=1.0\textwidth]{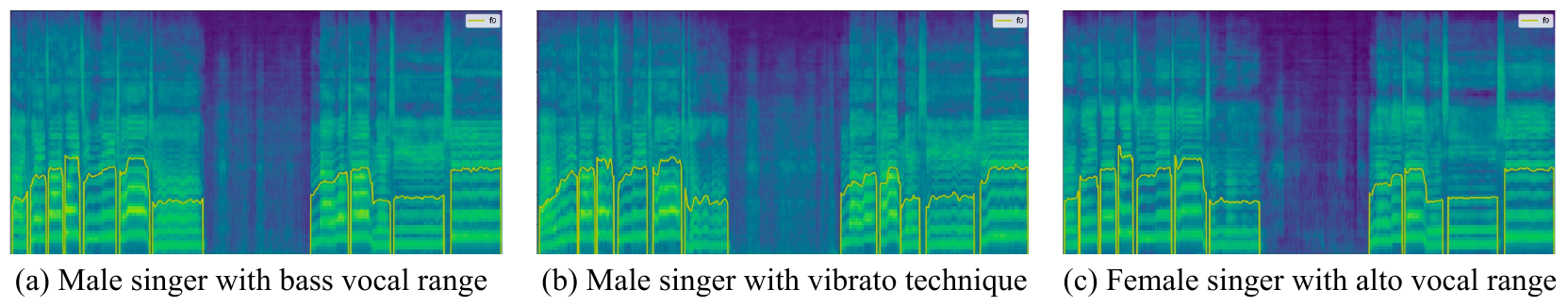}
\caption{
visualizations of style control. 
Figure (b) shows more F0 fluctuation than (a), highlighting vibrato. 
Figure (c) exhibits higher formants and richer high-frequency details than (a), reflecting different singers' identities.
}
\label{fig: sc}
\end{figure*}

\paragraph{Style Transfer.}

Table \ref{tab: base} presents the performance of TCSinger 2 compared to baseline models in the zero-shot style transfer task.
For the parallel experiments, we randomly select samples with unseen singers from the test set as target voices and use different utterances from the same singers to form prompts. Additionally, we utilize unseen test data with different lyric languages (such as English and Chinese) as prompts and targets for inference.
As shown in the results, TCSinger 2 demonstrates exceptional synthesis quality in both parallel and cross-lingual experiments, evidenced by the highest MOS-Q and the lowest FFE. This can be attributed to the naturalness introduced by the BBC Encoder, as well as the quality improvements from the linguistic-MOE and F0 supervision within the Flow-based Custom Transformer.
Moreover, TCSinger 2 also excels in singer similarity, as reflected by the highest MOS-S and Cos values. 
This highlights the effectiveness of the Custom Audio Encoder in capturing rich style information in the audio representation, as well as the improved style modeling enabled by the Stylistic-MOE.
Upon listening to the demos, it is evident that our model effectively transfers various aspects of singing style, including timbre, singing method, emotion, accent, and other nuanced elements from audio prompts.

\paragraph{Style Control.}
Table \ref{tab: sc} presents the experimental results for style control using natural language textual prompts.
We add a cross-attention model to the baseline models to handle the textual prompt.
In the parallel experiments, we randomly select unseen audio from the test set, using the ground truth (GT) textual prompts as the target. For the non-parallel experiments, multi-level styles are randomly assigned in a manner that is appropriate for the context. These styles include global timbre (such as the singer's gender and vocal range), singing method (e.g., bel canto and pop), emotion (e.g., happy and sad), and segment-level or word-level techniques (such as mixed voice, falsetto, breathy, vibrato, glissando, and pharyngeal).
As shown in the results, TCSinger 2 outperforms the baseline models in both the highest synthesis quality (MOS-Q and FFE) and style controllability (MOS-C) in both parallel and non-parallel experiments. This reflects the quality improvements brought by the BBC Encoder, Lingual-MOE, and F0 supervision, as well as the enhanced style control achieved through self-attention and the Stylistic-MOE.
These results demonstrate that, in addition to style transfer, TCSinger 2 also performs well in style control. 

Figure \ref{fig: sc} shows that we can effectively control diverse styles. 
Figure (b) demonstrates the vibrato technique, appearing as regular oscillations in F0. 
Figure (c), representing a female alto singer, exhibits higher formant frequencies, resulting in a generally upward-shifted energy distribution and richer high-frequency harmonic content compared to the male bass singer. 
Our demos show that TCSinger 2 can control multi-level styles effectively.

\begin{table*}[ht]
\centering
\small
\scalebox{1}{
\begin{tabular}{lcccccc}
\toprule
{\bfseries{Method}} & {FFE $\downarrow$} & {Cos $\uparrow$} & {MOS-Q $\uparrow$} & {MOS-S $\uparrow$} \\
\midrule
GT & - & - & 4.53 $\pm$ 0.11 & - \\
GT (vocoder) & 0.06 & 0.93 & 4.21 $\pm$ 0.08 & 4.20 $\pm$ 0.13 \\
\midrule
StyleTTS 2 & 0.41 & 0.71 & 3.60 $\pm$ 0.15 & 3.52 $\pm$ 0.10\\
CosyVoice & 0.39 & 0.79 & 3.66 $\pm$ 0.09 & 3.65 $\pm$ 0.14 \\
VISinger 2 & 0.32 & 0.75 & 3.72 $\pm$ 0.18 & 3.59 $\pm$ 0.07 \\
TCSinger & 0.28 & 0.82 & 3.89 $\pm$ 0.06 & 3.84 $\pm$ 0.16\\
\midrule
TCSinger 2 (ours) & \bf 0.24 & \bf 0.89 & \bf 3.97 $\pm$ 0.12 & \bf 3.96 $\pm$ 0.09\\
\bottomrule
\end{tabular}}
\caption{
Zero-shot speech-to-singing style transfer performance.
}
\label{tab: sts}
\end{table*}

\begin{table*}[t]
\centering
\small
\begin{tabular}{lcccc}
\toprule
\multirow{2}{*}{\bfseries{Setting}} & \multicolumn{2}{c}{\bfseries{Style Transfer}} & \multicolumn{2}{c}{\bfseries{Style Control}}\\ \cmidrule(lr){2-3} \cmidrule(lr){4-5}
  & CMOS-Q & CMOS-S & CMOS-Q & CMOS-C \\
\midrule
TCSinger 2 & 0.00 & 0.00 & 0.00 & 0.00 \\
\midrule
w/o BBC Encoder & -0.36 & -0.23 & -0.39 & -0.26 \\
w/o CAE  & -0.21  & -0.37 & -0.19 & -0.41\\
w/o F0 Supervision & -0.33  & -0.24 & -0.31 & -0.27\\
w/o CFG & -0.26  & -0.22 & -0.25 & -0.31\\
w/o Cus-MOE  &  -0.31 & -0.32 & -0.38 & -0.35\\
w/o Lingual-MOE  &  -0.29 & -0.17 & -0.32 & -0.21\\
w/o Stylistic-MOE  &  -0.21 & -0.26 & -0.23 & -0.33\\
\bottomrule   
\end{tabular}
\caption{
Style transfer and style control comparisons for the ablation study.
CAE denotes Custom Audio Encoder.
}
\label{tab: abl}
\end{table*}

\paragraph{Speech-to-Singing.}
We also conduct experiments on speech-to-singing style transfer. 
We randomly select unseen singers from the test set as target samples and different speech samples from the same singers to form the prompts. 
As shown in Table \ref{tab: sts}, both the synthesis quality (MOS-Q and FFE) and singer similarity (MOS-S and Cos) of TCSinger 2 outperform those of the baseline models. 
This demonstrates the ability of our Custom Audio Encoder to extend to a broader range of applications, enabling users who cannot sing to customize their singing voice using only speech prompts.

\subsection{Ablation Study}

As depicted in Table \ref{tab: abl}, we conduct ablation studies on style transfer and style control to demonstrate the efficacy of various designs within TCSinger 2. 
We use CMOS-Q to test variations in synthesis quality, CMOS-S to measure changes in singer similarity, and CMOS-C to evaluate differences in style controllability.
We first test the effect of removing the masking process from the BBC Encoder, and observe that CMOS-Q drops significantly, indicating a substantial impact on the naturalness of the generated results.
Then, we also test replacing the Custom Audio Encoder with a standard VAE encoder for both speech and singing prompts, which leads to a decline in CMOS-S and CMOS-C, showing that it negatively affects style modeling.

Next, we test several designs in the Flow-Based Transformer. When we do not use F0 supervision, we observe a decline in all metrics—CMOS-Q, CMOS-S, and CMOS-C, consistent with our understanding of the important role pitch modeling plays in SVS.
We also test the scenario without using the CFG strategy and find that CMOS-Q, CMOS-C, and CMOS-S decrease significantly, while CMOS-Q only shows a slight decline. 
This demonstrates the contribution of the CFG strategy to improving style transfer, style control, and synthesis quality.

Finally, we test the performance of Cus-MOE and the scenario where each expert group is replaced with a standard FFN. We observe that Cus-MOE impacts all aspects, while Lingual-MOE primarily affects the quality of multilingual SVS (CMOS-Q), and Stylistic-MOE mainly influences style transfer and style control (CMOS-S and CMOS-C).
These experiments collectively demonstrate the effectiveness of the various designs in TCSinger 2 for multi-task multilingual zero-shot SVS with style transfer and style control. 
For more extensive experiments, please refer to Appendix \ref{sec: appendix5}.

\section{Conclusion}

In this paper, we present TCSinger 2, a multilingual, multi-task, zero-shot singing voice synthesis model with advanced style transfer and style control capabilities based on various prompts.
To ensure smooth and robust phoneme/note transitions, we introduce the Blurred Boundary Content Encoder, which applies flexible boundary masking on phoneme and note boundaries for seamless transitions.
For aligned representations across singing, speech, and textual prompts, we propose the Custom Audio Encoder using contrastive learning, broadening the model’s applicability to a wide range of tasks.
Moreover, we also introduce the Flow-based Custom Transformer to stably and fastly generate high-quality, highly controllable singing voices. 
The model employs Cus-MOE and F0 supervision to optimize synthesis quality and style modeling.
Our experimental results show that TCSinger 2 outperforms other baseline models in synthesis quality, singer similarity, and style controllability across various related tasks, including zero-shot style transfer, cross-lingual style transfer, multi-level style control, and STS style transfer.

\section{Limitations}

Our method has two main limitations.
First, it still relies on manually labeled styles, which introduces errors in style annotations and is constrained by the high cost of dataset labeling. Future work will explore the use of automatic labeling tools to expand datasets at a lower cost, thereby improving the SVS model's generalization ability.
Second, although our model accelerates inference speed through the flow-matching structure, the generation speed still does not meet higher industrial demands. 
In future work, we will investigate streaming generation methods to reduce latency.

\section{Ethics Statement}

TCSinger 2, with its ability to adapt and manipulate various singing styles, carries the potential for misuse in the dubbing of entertainment content, which could lead to violations of singers' intellectual property rights.
Moreover, the model’s capability to control styles using diverse prompts introduces the risk of unfair competition and the possible displacement of professionals in the singing industry.
To address these concerns, we plan to impose strict regulations on the model’s usage to prevent unethical and unauthorized applications.
Additionally, we will investigate methods like vocal watermarking to ensure the protection of individual privacy.

\section*{Acknowledgements}

This work was supported by National Natural Science Foundation of China under Grant No. 62222211 and National Natural Science Foundation of China under Grant No.U24A20326.

\bibliography{custom}

\begin{thebibliography}{44}
\providecommand{\natexlab}[1]{#1}

\bibitem[{Baevski et~al.(2020)Baevski, Zhou, Mohamed, and Auli}]{baevski2020wav2vec}
Alexei Baevski, Yuhao Zhou, Abdelrahman Mohamed, and Michael Auli. 2020.
\newblock wav2vec 2.0: A framework for self-supervised learning of speech representations.
\newblock \emph{Advances in neural information processing systems}, 33:12449--12460.

\bibitem[{Chen et~al.(2022)Chen, Wang, Chen, Wu, Liu, Chen, Li, Kanda, Yoshioka, Xiao et~al.}]{chen2022wavlm}
Sanyuan Chen, Chengyi Wang, Zhengyang Chen, Yu~Wu, Shujie Liu, Zhuo Chen, Jinyu Li, Naoyuki Kanda, Takuya Yoshioka, Xiong Xiao, et~al. 2022.
\newblock Wavlm: Large-scale self-supervised pre-training for full stack speech processing.
\newblock \emph{IEEE Journal of Selected Topics in Signal Processing}, 16(6):1505--1518.

\bibitem[{Cho et~al.(2022)Cho, Tsao, Wang, and Liu}]{cho2022mandarin}
Yin-Ping Cho, Yu~Tsao, Hsin-Min Wang, and Yi-Wen Liu. 2022.
\newblock Mandarin singing voice synthesis with denoising diffusion probabilistic wasserstein gan.
\newblock In \emph{2022 Asia-Pacific Signal and Information Processing Association Annual Summit and Conference (APSIPA ASC)}, pages 1956--1963. IEEE.

\bibitem[{Choi et~al.(2020)Choi, Han, Kim, and Ha}]{choi2020attentron}
Seungwoo Choi, Seungju Han, Dongyoung Kim, and Sungjoo Ha. 2020.
\newblock Attentron: Few-shot text-to-speech utilizing attention-based variable-length embedding.
\newblock \emph{arXiv preprint arXiv:2005.08484}.

\bibitem[{Choi and Nam(2022)}]{choi2022melody}
Soonbeom Choi and Juhan Nam. 2022.
\newblock A melody-unsupervision model for singing voice synthesis.
\newblock In \emph{ICASSP 2022-2022 IEEE International Conference on Acoustics, Speech and Signal Processing (ICASSP)}, pages 7242--7246. IEEE.

\bibitem[{Chung et~al.(2024)Chung, Hou, Longpre, Zoph, Tay, Fedus, Li, Wang, Dehghani, Brahma et~al.}]{chung2024scaling}
Hyung~Won Chung, Le~Hou, Shayne Longpre, Barret Zoph, Yi~Tay, William Fedus, Yunxuan Li, Xuezhi Wang, Mostafa Dehghani, Siddhartha Brahma, et~al. 2024.
\newblock Scaling instruction-finetuned language models.
\newblock \emph{Journal of Machine Learning Research}, 25(70):1--53.

\bibitem[{Cui et~al.(2024)Cui, Gu, Weng, Zhang, Chen, and Dai}]{cui2024sifisinger}
Jianwei Cui, Yu~Gu, Chao Weng, Jie Zhang, Liping Chen, and Lirong Dai. 2024.
\newblock Sifisinger: A high-fidelity end-to-end singing voice synthesizer based on source-filter model.
\newblock In \emph{ICASSP 2024-2024 IEEE International Conference on Acoustics, Speech and Signal Processing (ICASSP)}, pages 11126--11130. IEEE.

\bibitem[{Du et~al.(2024)Du, Chen, Zhang, Hu, Lu, Yang, Hu, Zheng, Gu, Ma et~al.}]{du2024cosyvoice}
Zhihao Du, Qian Chen, Shiliang Zhang, Kai Hu, Heng Lu, Yexin Yang, Hangrui Hu, Siqi Zheng, Yue Gu, Ziyang Ma, et~al. 2024.
\newblock Cosyvoice: A scalable multilingual zero-shot text-to-speech synthesizer based on supervised semantic tokens.
\newblock \emph{arXiv preprint arXiv:2407.05407}.

\bibitem[{Fedus et~al.(2022)Fedus, Zoph, and Shazeer}]{fedus2022switch}
William Fedus, Barret Zoph, and Noam Shazeer. 2022.
\newblock Switch transformers: Scaling to trillion parameter models with simple and efficient sparsity.
\newblock \emph{Journal of Machine Learning Research}, 23(120):1--39.

\bibitem[{Guo et~al.(2025)Guo, Zhang, Pan, Huang, Tang, Li, Hong, Wang, and Zhao}]{guo2025techsinger}
Wenxiang Guo, Yu~Zhang, Changhao Pan, Rongjie Huang, Li~Tang, Ruiqi Li, Zhiqing Hong, Yongqi Wang, and Zhou Zhao. 2025.
\newblock Techsinger: Technique controllable multilingual singing voice synthesis via flow matching.
\newblock \emph{arXiv preprint arXiv:2502.12572}.

\bibitem[{Huang et~al.(2021)Huang, Chen, Ren, Liu, Cui, and Zhao}]{huang2021multi}
Rongjie Huang, Feiyang Chen, Yi~Ren, Jinglin Liu, Chenye Cui, and Zhou Zhao. 2021.
\newblock Multi-singer: Fast multi-singer singing voice vocoder with a large-scale corpus.
\newblock In \emph{Proceedings of the 29th ACM International Conference on Multimedia}, pages 3945--3954.

\bibitem[{Jiang et~al.(2025)Jiang, Ren, Li, Ji, Zhang, Ye, Zhang, Jionghao, Yang, Zuo et~al.}]{jiang2025megatts}
Ziyue Jiang, Yi~Ren, Ruiqi Li, Shengpeng Ji, Boyang Zhang, Zhenhui Ye, Chen Zhang, Bai Jionghao, Xiaoda Yang, Jialong Zuo, et~al. 2025.
\newblock Megatts 3: Sparse alignment enhanced latent diffusion transformer for zero-shot speech synthesis.
\newblock \emph{arXiv preprint arXiv:2502.18924}.

\bibitem[{Kim et~al.(2023)Kim, Kim, Jun, and Kim}]{kim2023muse}
Sungjae Kim, Yewon Kim, Jewoo Jun, and Injung Kim. 2023.
\newblock Muse-svs: Multi-singer emotional singing voice synthesizer that controls emotional intensity.
\newblock \emph{IEEE/ACM Transactions on Audio, Speech, and Language Processing}.

\bibitem[{Kim et~al.(2022)Kim, Kang, and Lee}]{kim2022adversarial}
Tae-Woo Kim, Min-Su Kang, and Gyeong-Hoon Lee. 2022.
\newblock Adversarial multi-task learning for disentangling timbre and pitch in singing voice synthesis.
\newblock \emph{arXiv preprint arXiv:2206.11558}.

\bibitem[{Kingma and Welling(2013)}]{kingma2013auto}
Diederik~P Kingma and Max Welling. 2013.
\newblock Auto-encoding variational bayes.
\newblock \emph{arXiv preprint arXiv:1312.6114}.

\bibitem[{Kong et~al.(2020)Kong, Kim, and Bae}]{kong2020hifi}
Jungil Kong, Jaehyeon Kim, and Jaekyoung Bae. 2020.
\newblock Hifi-gan: Generative adversarial networks for efficient and high fidelity speech synthesis.
\newblock \emph{Advances in neural information processing systems}, 33:17022--17033.

\bibitem[{Kumar et~al.(2021)Kumar, Goel, Narang, and Lall}]{kumar2021normalization}
Neeraj Kumar, Srishti Goel, Ankur Narang, and Brejesh Lall. 2021.
\newblock Normalization driven zero-shot multi-speaker speech synthesis.
\newblock In \emph{Interspeech}, pages 1354--1358.

\bibitem[{Li et~al.(2024{\natexlab{a}})Li, Zhang, Wang, Hong, Huang, and Zhao}]{li2024robust}
Ruiqi Li, Yu~Zhang, Yongqi Wang, Zhiqing Hong, Rongjie Huang, and Zhou Zhao. 2024{\natexlab{a}}.
\newblock Robust singing voice transcription serves synthesis.
\newblock \emph{arXiv preprint arXiv:2405.09940}.

\bibitem[{Li et~al.(2022)Li, Han, and Mesgarani}]{li2022styletts}
Yinghao~Aaron Li, Cong Han, and Nima Mesgarani. 2022.
\newblock Styletts: A style-based generative model for natural and diverse text-to-speech synthesis.
\newblock \emph{arXiv preprint arXiv:2205.15439}.

\bibitem[{Li et~al.(2024{\natexlab{b}})Li, Han, Raghavan, Mischler, and Mesgarani}]{li2024styletts}
Yinghao~Aaron Li, Cong Han, Vinay Raghavan, Gavin Mischler, and Nima Mesgarani. 2024{\natexlab{b}}.
\newblock Styletts 2: Towards human-level text-to-speech through style diffusion and adversarial training with large speech language models.
\newblock \emph{Advances in Neural Information Processing Systems}, 36.

\bibitem[{Liu et~al.(2022{\natexlab{a}})Liu, Li, Ren, Zhu, and Zhao}]{liu2022learning}
Jinglin Liu, Chengxi Li, Yi~Ren, Zhiying Zhu, and Zhou Zhao. 2022{\natexlab{a}}.
\newblock Learning the beauty in songs: Neural singing voice beautifier.
\newblock \emph{arXiv preprint arXiv:2202.13277}.

\bibitem[{Liu et~al.(2022{\natexlab{b}})Liu, Gong et~al.}]{liu2022flow}
Xingchao Liu, Chengyue Gong, et~al. 2022{\natexlab{b}}.
\newblock Flow straight and fast: Learning to generate and transfer data with rectified flow.
\newblock In \emph{The Eleventh International Conference on Learning Representations}.

\bibitem[{Mao et~al.(2017)Mao, Li, Xie, Lau, Wang, and Paul~Smolley}]{mao2017least}
Xudong Mao, Qing Li, Haoran Xie, Raymond~YK Lau, Zhen Wang, and Stephen Paul~Smolley. 2017.
\newblock Least squares generative adversarial networks.
\newblock In \emph{Proceedings of the IEEE international conference on computer vision}, pages 2794--2802.

\bibitem[{McAuliffe et~al.(2017)McAuliffe, Socolof, Mihuc, Wagner, and Sonderegger}]{mcauliffe2017montreal}
Michael McAuliffe, Michaela Socolof, Sarah Mihuc, Michael Wagner, and Morgan Sonderegger. 2017.
\newblock Montreal forced aligner: Trainable text-speech alignment using kaldi.
\newblock In \emph{Interspeech}, volume 2017, pages 498--502.

\bibitem[{Nie et~al.(2021)Nie, Miao, Cao, Ma, Liu, Xue, Miao, Liu, Yang, and Cui}]{nie2021evomoe}
Xiaonan Nie, Xupeng Miao, Shijie Cao, Lingxiao Ma, Qibin Liu, Jilong Xue, Youshan Miao, Yi~Liu, Zhi Yang, and Bin Cui. 2021.
\newblock Evomoe: An evolutional mixture-of-experts training framework via dense-to-sparse gate.
\newblock \emph{arXiv preprint arXiv:2112.14397}.

\bibitem[{Peebles and Xie(2023)}]{peebles2023scalable}
William Peebles and Saining Xie. 2023.
\newblock Scalable diffusion models with transformers.
\newblock In \emph{Proceedings of the IEEE/CVF International Conference on Computer Vision}, pages 4195--4205.

\bibitem[{Radford et~al.(2021)Radford, Kim, Hallacy, Ramesh, Goh, Agarwal, Sastry, Askell, Mishkin, Clark et~al.}]{radford2021learning}
Alec Radford, Jong~Wook Kim, Chris Hallacy, Aditya Ramesh, Gabriel Goh, Sandhini Agarwal, Girish Sastry, Amanda Askell, Pamela Mishkin, Jack Clark, et~al. 2021.
\newblock Learning transferable visual models from natural language supervision.
\newblock In \emph{International conference on machine learning}, pages 8748--8763. PMLR.

\bibitem[{Skerry-Ryan et~al.(2018)Skerry-Ryan, Battenberg, Xiao, Wang, Stanton, Shor, Weiss, Clark, and Saurous}]{skerry2018towards}
RJ~Skerry-Ryan, Eric Battenberg, Ying Xiao, Yuxuan Wang, Daisy Stanton, Joel Shor, Ron Weiss, Rob Clark, and Rif~A Saurous. 2018.
\newblock Towards end-to-end prosody transfer for expressive speech synthesis with tacotron.
\newblock In \emph{international conference on machine learning}, pages 4693--4702. PMLR.

\bibitem[{Su et~al.(2024)Su, Ahmed, Lu, Pan, Bo, and Liu}]{su2024roformer}
Jianlin Su, Murtadha Ahmed, Yu~Lu, Shengfeng Pan, Wen Bo, and Yunfeng Liu. 2024.
\newblock Roformer: Enhanced transformer with rotary position embedding.
\newblock \emph{Neurocomputing}, 568:127063.

\bibitem[{Wagner and Watson(2010)}]{wagner2010experimental}
Michael Wagner and Duane~G Watson. 2010.
\newblock Experimental and theoretical advances in prosody: A review.
\newblock \emph{Language and cognitive processes}, 25(7-9):905--945.

\bibitem[{Wang et~al.(2024)Wang, Hu, Huang, Hong, Li, Liu, You, Jin, and Zhao}]{wang2024prompt}
Yongqi Wang, Ruofan Hu, Rongjie Huang, Zhiqing Hong, Ruiqi Li, Wenrui Liu, Fuming You, Tao Jin, and Zhou Zhao. 2024.
\newblock Prompt-singer: Controllable singing-voice-synthesis with natural language prompt.
\newblock \emph{arXiv preprint arXiv:2403.11780}.

\bibitem[{Wang et~al.(2022)Wang, Wang, Zhu, Wu, Li, Xue, Zhang, Xie, and Bi}]{wang2022opencpop}
Yu~Wang, Xinsheng Wang, Pengcheng Zhu, Jie Wu, Hanzhao Li, Heyang Xue, Yongmao Zhang, Lei Xie, and Mengxiao Bi. 2022.
\newblock Opencpop: A high-quality open source chinese popular song corpus for singing voice synthesis.
\newblock \emph{arXiv preprint arXiv:2201.07429}.

\bibitem[{Za{\i}di et~al.(2021)Za{\i}di, Seut{\'e}, Niekerk, and Carbonneau}]{zaidi2021daft}
Julian Za{\i}di, Hugo Seut{\'e}, BV~Niekerk, and M~Carbonneau. 2021.
\newblock Daft-exprt: Robust prosody transfer across speakers for expressive speech synthesis.
\newblock \emph{arXiv preprint arXiv:2108.02271}.

\bibitem[{Zhang and Sennrich(2019)}]{zhang2019root}
Biao Zhang and Rico Sennrich. 2019.
\newblock Root mean square layer normalization.
\newblock \emph{Advances in Neural Information Processing Systems}, 32.

\bibitem[{Zhang et~al.(2022{\natexlab{a}})Zhang, Li, Wang, Deng, Liu, Ren, He, Huang, Zhu, Chen et~al.}]{zhang2022m4singer}
Lichao Zhang, Ruiqi Li, Shoutong Wang, Liqun Deng, Jinglin Liu, Yi~Ren, Jinzheng He, Rongjie Huang, Jieming Zhu, Xiao Chen, et~al. 2022{\natexlab{a}}.
\newblock M4singer: A multi-style, multi-singer and musical score provided mandarin singing corpus.
\newblock \emph{Advances in Neural Information Processing Systems}, 35:6914--6926.

\bibitem[{Zhang et~al.(2025{\natexlab{a}})Zhang, Chen, Liu, Xi, Huo, Liu, and Wu}]{zhang2025sgw}
Ruiyuan Zhang, Yuyao Chen, Jiaxiang Liu, Dianbing Xi, Yuchi Huo, Jie Liu, and Chao Wu. 2025{\natexlab{a}}.
\newblock Sgw-based multi-task learning in vision tasks.
\newblock In \emph{Asian Conference on Computer Vision}, pages 124--141. Springer.

\bibitem[{Zhang et~al.(2024{\natexlab{a}})Zhang, Liu, Li, Dong, Fu, and Wu}]{zhang2024scalable}
Ruiyuan Zhang, Jiaxiang Liu, Zexi Li, Hao Dong, Jie Fu, and Chao Wu. 2024{\natexlab{a}}.
\newblock Scalable geometric fracture assembly via co-creation space among assemblers.
\newblock In \emph{Proceedings of the AAAI Conference on Artificial Intelligence}, volume~38, pages 7269--7277.

\bibitem[{Zhang et~al.(2025{\natexlab{b}})Zhang, Wang, Liu, Zhang, Huo, and Wu}]{zhang2025leveraging}
Ruiyuan Zhang, Qi~Wang, Jiaxiang Liu, Yu~Zhang, Yuchi Huo, and Chao Wu. 2025{\natexlab{b}}.
\newblock Leveraging pretrained diffusion models for zero-shot part assembly.
\newblock \emph{arXiv preprint arXiv:2505.00426}.

\bibitem[{Zhang et~al.(2022{\natexlab{b}})Zhang, Cong, Xue, Xie, Zhu, and Bi}]{zhang2022visinger}
Yongmao Zhang, Jian Cong, Heyang Xue, Lei Xie, Pengcheng Zhu, and Mengxiao Bi. 2022{\natexlab{b}}.
\newblock Visinger: Variational inference with adversarial learning for end-to-end singing voice synthesis.
\newblock In \emph{ICASSP 2022-2022 IEEE International Conference on Acoustics, Speech and Signal Processing (ICASSP)}, pages 7237--7241. IEEE.

\bibitem[{Zhang et~al.(2025{\natexlab{c}})Zhang, Guo, Pan, Zhu, Jin, and Zhao}]{zhang2025isdrama}
Yu~Zhang, Wenxiang Guo, Changhao Pan, Zhiyuan Zhu, Tao Jin, and Zhou Zhao. 2025{\natexlab{c}}.
\newblock Isdrama: Immersive spatial drama generation through multimodal prompting.
\newblock \emph{arXiv preprint arXiv:2504.20630}.

\bibitem[{Zhang et~al.(2025{\natexlab{d}})Zhang, Guo, Pan, Zhu, Li, Lu, Huang, Zhang, Hong, Jiang et~al.}]{zhang2025versatile}
Yu~Zhang, Wenxiang Guo, Changhao Pan, Zhiyuan Zhu, Ruiqi Li, Jingyu Lu, Rongjie Huang, Ruiyuan Zhang, Zhiqing Hong, Ziyue Jiang, et~al. 2025{\natexlab{d}}.
\newblock Versatile framework for song generation with prompt-based control.
\newblock \emph{arXiv preprint arXiv:2504.19062}.

\bibitem[{Zhang et~al.(2024{\natexlab{b}})Zhang, Huang, Li, He, Xia, Chen, Duan, Huai, and Zhao}]{zhang2024stylesinger}
Yu~Zhang, Rongjie Huang, Ruiqi Li, JinZheng He, Yan Xia, Feiyang Chen, Xinyu Duan, Baoxing Huai, and Zhou Zhao. 2024{\natexlab{b}}.
\newblock Stylesinger: Style transfer for out-of-domain singing voice synthesis.
\newblock In \emph{Proceedings of the AAAI Conference on Artificial Intelligence}, volume~38, pages 19597--19605.

\bibitem[{Zhang et~al.(2024{\natexlab{c}})Zhang, Jiang, Li, Pan, He, Huang, Wang, and Zhao}]{zhang2024tcsinger}
Yu~Zhang, Ziyue Jiang, Ruiqi Li, Changhao Pan, Jinzheng He, Rongjie Huang, Chuxin Wang, and Zhou Zhao. 2024{\natexlab{c}}.
\newblock Tcsinger: Zero-shot singing voice synthesis with style transfer and multi-level style control.
\newblock In \emph{Proceedings of the 2024 Conference on Empirical Methods in Natural Language Processing}, pages 1960--1975.

\bibitem[{Zhang et~al.(2024{\natexlab{d}})Zhang, Pan, Guo, Li, Zhu, Wang, Xu, Lu, Hong, Wang et~al.}]{zhang2024gtsinger}
Yu~Zhang, Changhao Pan, Wenxiang Guo, Ruiqi Li, Zhiyuan Zhu, Jialei Wang, Wenhao Xu, Jingyu Lu, Zhiqing Hong, Chuxin Wang, et~al. 2024{\natexlab{d}}.
\newblock Gtsinger: A global multi-technique singing corpus with realistic music scores for all singing tasks.
\newblock \emph{arXiv preprint arXiv:2409.13832}.

\end{thebibliography}

\newpage
\appendix

\section{Details of Models}
\label{sec: appendix1}

\subsection{Architecture Details}
\label{sec: appendix1arch}

For the custom audio encoder and decoder, we adopt a Variational Autoencoder (VAE) architecture \citep{kingma2013auto}. The Mel-spectrograms are derived from waveforms sampled at 48 kHz, with a 1024 window size, a 256 hop size, and 80 Mel bins. HiFi-GAN \citep{kong2020hifi} is utilized as the vocoder to synthesize waveforms from the Mel-spectrograms.
The model architecture consists of three layers for both the encoder and decoder, with a hidden size of 384 and a Conv1D kernel size of five. The mel-spectrogram, with dimensions $B, 80, T$, is compressed to $B, 20, T/8$, facilitating further processing by the Transformer. 
Textual prompts are encoded with FLAN-T5-large \citep{chung2024scaling}.
During training, fixed-length batches containing 2000 mel-spectrogram frames are used. The Adam optimizer is employed with a learning rate of $1 \times 10^{-4}$, $\beta_1 = 0.9$, $\beta_2 = 0.999$, and a warm-up period of 10K steps.

For the Flow-based Custom Transformer, we utilize four Transformer blocks as the vector field estimator. Each transformer layer uses a hidden size of 768 and eight attention heads. The Cus-MoE architecture includes four experts per expert group. The total number of parameters is 105 million. Flow-matching during training uses 1,000 timesteps, while inference uses 25 timesteps with the Euler ODE solver. During training, we use eight NVIDIA RTX-4090 GPUs, with a batch size of 12K frames per GPU, for 100K steps. 
The Adam optimizer is applied with a learning rate of $5 \times 10^{-5}$, $\beta_1 = 0.9$, $\beta_2 = 0.999$, and 10K warm-up steps.

\subsection{Custom Audio Encoder}
\label{sec: appendix1vae}

Multi-level styles encompass global-level singing methods (such as bel canto) and emotional elements (such as happy or sad). The system incorporates segment-level or word-level techniques (such as mixed voice and falsetto). 
It also accounts for natural elements influenced by personal habits, such as accent, pronunciation, and transitions. 
Audio prompts (either singing or speech) enable the target singing voice to learn and mimic all of these styles, while textual prompts offer the flexibility to control both global and word-level styles.
We design three types of contrasts: (1) same content, different styles; (2) similar styles, different content; and (3) different styles and different content.
For the first type of contrast, we use different multi-level styles for the same song. For the second type, we apply similar labels but different song contents (e.g., different phrases from the same song).
For speech and singing contrasts in the first type, we use different singers (speakers) performing the same lyrics, which introduces various natural elements. For the second type, we use the same singer for different parts of the song (e.g., different phrases of the same song).
In the comparison between textual prompts and speech, we contrast the global styles of the lyrics in the speech to those in the textual prompts.

\subsection{Flow-based Custom Transformer}
\label{sec: appendix1flow}
In generative models, the true data distribution is denoted by $q(x_1)$, which can be sampled but lacks an explicit probability density function. A probabilistic path $p_t(x_t)$ links the standard Gaussian distribution $x_0 \sim p_0(x)$ to the actual data distribution $x_1 \sim p_1(x)$. The flow-matching technique \citep{liu2022flow} models this transformation by solving the ordinary differential equation (ODE):
\begin{equation}
\begin{aligned}
&\mathrm{d}x = u(x, t) \, \mathrm{d}t, \quad t \in [0, 1],
\end{aligned}
\end{equation}
where $u(x, t)$ represents the target vector field and $t$ is the time parameter. With access to $u(x, t)$, realistic data can be generated by reversing the flow. To estimate $u(x, t)$, a vector field estimator $v(x, t; \theta)$ is employed, and the flow-matching objective is:
\begin{equation}
\begin{aligned}
&\mathcal{L}_{\mathrm{FM}}(\theta) &= \mathbb{E}_{t, p_t(x)} \left\| v(x, t; \theta) - u(x, t) \right\|^2.
\end{aligned}
\end{equation}
For conditional data, the objective is modified as:
\begin{equation}
\begin{aligned}
&\mathcal{L}_{\mathrm{CFM}}(\theta) = \mathbb{E}_{t, p_1(x_1), p_t(x | x_1)} \\ 
&\left\| v(x, t | C; \theta)- u(x, t | x_1, C) \right\|^2.
\end{aligned}
\end{equation}
Flow-matching directly transforms Gaussian noise into real data by linearly interpolating between $x_0$ and $x_1$ to generate samples at a given time $t$:
\begin{equation}
\begin{aligned}
&x_t = (1 - t) x_0 + t x_1.
\end{aligned}
\end{equation}
Thus, the conditional vector field becomes $u(x, t | x_1, C) = x_1 - x_0$, and the rectified flow-matching (RFM) loss is:
\begin{equation}
\begin{aligned}
&\left\| v(x, t | C; \theta) - (x_1 - x_0) \right\|^2.
\end{aligned}
\end{equation}
Once the vector field is accurately estimated, realistic data can be generated by solving the ODE using an Euler solver:
\begin{equation}
\begin{aligned}
&x_{t + \epsilon} &= x + \epsilon v(x, t | C; \theta),
\end{aligned}
\end{equation}
where $\epsilon$ represents the step size. 
Flow-matching models typically require hundreds or even thousands of training iterations. However, by utilizing linear interpolation, this number can be reduced to 25 steps or fewer during inference, resulting in significant computational efficiency improvements. This interpolation guarantees smooth transitions from noise to data, generating high-quality outputs without artifacts and ensuring consistency across different conditions.
Compared to simple transformer or diffusion methods \citep{zhang2025leveraging,zhang2024scalable,zhang2025sgw}, the flow-based transformer is more effective.

To enhance training stability and prevent numerical instability, we apply RMSNorm \citep{zhang2019root}.
The global embedding $z_g$ is computed by averaging the audio prompt $z_{pa}$ or textual prompt embedding $z_{pt}$ over the temporal dimension, with the time step embedding $z_t$ added. This global embedding is processed through a global adaptor using adaptive layer normalization (AdaLN) \citep{peebles2023scalable} to ensure consistent style. The AdaLN operation is defined as:
\begin{equation}
\begin{aligned}
&AdaLN(h, c) &= \gamma_c \times \text{LayerNorm}(h) + \beta_c,
\end{aligned}
\end{equation}
where $h$ represents the hidden representation, and the batch normalization scale $\gamma$ is initialized to zero \citep{peebles2023scalable}. 
Rotary positional embeddings (RoPE) \citep{su2024roformer} are employed to encode temporal positional information, improving the model’s capacity to capture dependencies across sequential frames.

\subsection{Cus-MOE}
\label{sec: appendix1moe}

Our routing mechanism leverages the dense-to-sparse Gumbel-Softmax technique \citep{nie2021evomoe} for efficient and adaptive expert selection. This method employs the Gumbel-Softmax trick to reparameterize categorical variables, making sampling differentiable and enabling dynamic routing.
For a given hidden state $h$, the routing score assigned to expert $i$, denoted as $g(h)_i$, is :
\begin{equation}
\begin{aligned}
&g(h)_i = \frac{\exp((h \cdot W_g + \zeta_i) / \tau)}{\sum_{j=1}^N \exp((h \cdot W_g + \zeta_j) / \tau)},
\end{aligned}
\end{equation}
where $W_g$ is the trainable gating weight, $\zeta$ is noise sampled from a Gumbel(0, 1) distribution, and $\tau$ represents the softmax temperature.
At the start of training, $\tau$ is set to a high value to encourage denser routing, allowing multiple experts to contribute to the processing of the same input. As training progresses, $\tau$ is gradually reduced, resulting in more selective routing with fewer experts involved. When $\tau$ approaches zero, the output distribution becomes nearly one-hot, with each token being assigned to the most relevant expert.
Following the approach outlined by \citep{nie2021evomoe}, we reduce $\tau$ from 2.0 to 0.3 during training, transitioning from dense to sparse routing. During inference, deterministic routing is employed, ensuring that only one expert is chosen for each token.

In our implementation, the regularization strength for the load balance loss is set to 0.1. The load-balancing mechanism promotes a more balanced distribution of tokens across experts, improving training efficiency by preventing underutilization or overload of specific experts. This routing strategy not only facilitates dynamic expert selection but also ensures an even distribution of computational resources.

\section{Details of Dataset}
\label{sec: appendix2}

\begin{table}[t]
\centering
\small
\begin{tabular}{lcccc}
\toprule
\bf Dataset & \bf Languages & \bf Singing/h & \bf Speech/h \\
\midrule
Opencpop & 1 & 5 & 0\\
M4Singer &1 & 30 & 0\\
OpenSinger & 1 & 85 & 0 \\
BuTFy & 1 & 8 & 10 \\
GTSinger& 9 & 80 & 16 \\
Extended & 5 & 50 & 5 \\
\midrule
\bfseries{Total/h} & 9 & 258 & 31\\
\bottomrule   
\end{tabular}
\caption{\label{tab: data}
Time distribution of our datasets.
}
\end{table}

The dataset for singing voices is quite limited. However, using the blurred boundary strategy, we expand our dataset by collecting 50 hours of clean singing voices and annotating them.
We also use several open-source singing datasets, including Opencpop \citep{wang2022opencpop} (Chinese, 1 singer, 5 hours of singing voices), M4Singer \citep{zhang2022m4singer} (Chinese, 20 singers, 30 hours of singing voices), OpenSinger \citep{huang2021multi} (Chinese, 93 singers, 85 hours of singing voices), PopBuTFy \citep{liu2022learning} (English, 20 singers, 18 hours of speech and singing voices), and GTSinger \citep{zhang2024gtsinger} (9 languages, 20 singers, 80 hours of singing and speech). 
We use all these datasets under the CC BY-NC-SA 4.0 license. 
All languages include Chinese, English, French, Spanish, German, Italian, Japanese, Korean, and Russian.
The time distribution of our datasets is listed in Table \ref{tab: data}.

For datasets without music scores and alignments, we use ROSVOT \citep{li2024robust} for coarse music score annotations and the Montreal Forced Aligner (MFA) \citep{mcauliffe2017montreal} for coarse alignment between lyrics and audio. 
Moreover, with the assistance of music experts, we manually annotate part of the singing data with multi-level style labels. 
We label the timbre of these data, including gender and vocal range.
We categorize songs as happy or sad based on emotion. 
For singing methods, we classify songs as bel canto or pop. These classifications are then combined into the final style class labels, which will serve as the global text prompts.
We also annotate segment-level and word-level techniques for these singing data. 
These techniques include mixed voice, falsetto, breathy, vibrato, glissando, and pharyngeal.
These technique labels form the segment-level and word-level text prompts.
All music experts and annotators we hire have musical backgrounds, and they are compensated at a rate of \$300 per hour. 
They have agreed to make their contributions available for research purposes.
Finally, we use GPT-4o to convert these labels into natural language textual prompts, like \textit{A female singer with an alto vocal range performing a happy pop song. She begins with a mixed voice in the first half of the song, showcasing a smooth and bright tone, before transitioning into falsetto for the second half, bringing an uplifting and energetic feel to the performance.} 

\section{Details of Evaluation}
\label{sec: appendix3}

\subsection{Subjective Evaluation}
\label{sec: appendix3sub}

\begin{figure*}[t]
\centering
\includegraphics[width=1.0\textwidth]{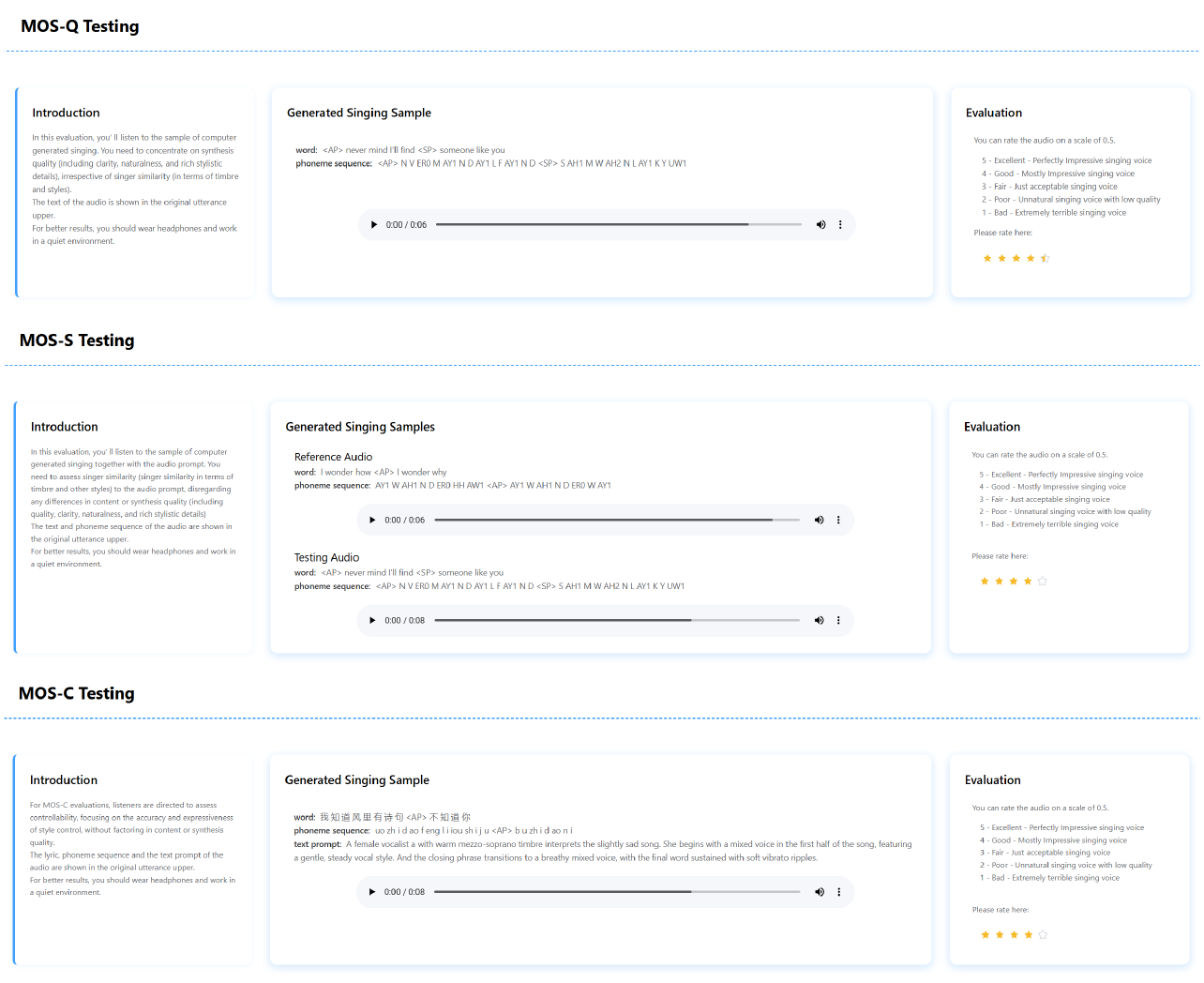}
\caption{
The instructions for our subjective evaluation on MOS.
}
\label{fig: eva}
\end{figure*}

For each evaluation task, we randomly select 40 pairs of sentences from our test set for subjective assessment. Each pair consists of an audio prompt or a textual prompt that defines styles, along with a synthesized singing voice. 
These pairs are presented to at least 10 professional listeners for review.
We utilize MOS (Mean Opinion Score) and CMOS (Comparative Mean Opinion Score) as the subjective evaluation metrics. 
In the MOS-Q and CMOS-Q evaluations, listeners are instructed to focus on the synthesis quality, including clarity, naturalness, and stylistic richness, without considering singer similarity (in terms of timbre and other styles). 
In contrast, for MOS-S and CMOS-S evaluations, listeners are asked to evaluate singer similarity, specifically the resemblance to the timbre and other styles of the audio prompt, while disregarding any differences in content or synthesis quality.
For MOS-C evaluations, listeners are directed to assess controllability, focusing on the accuracy and expressiveness of style control, without factoring in content or synthesis quality.
In all MOS-Q, MOS-S, and MOS-C evaluations, listeners rate the various singing voice samples on a Likert scale from 1 to 5. In the CMOS-Q and CMOS-S evaluations, listeners are tasked with comparing pairs of singing voices produced by different systems and expressing their preferences. The preference scale is as follows: 0 for no difference, 1 for a slight difference, and 2 for a significant difference.
It is important to note that all participants are fairly compensated for their time and effort. Each participant is paid \$10 per hour, resulting in a total expenditure of approximately \$300 for participant compensation. Participants are also informed that the results will be used for scientific research purposes.
The instruction screenshots are shown in Figure \ref{fig: eva}.

\subsection{Objective Evaluation}
\label{sec: appendix3obj}

To objectively assess the timbre similarity and synthesis quality of the test set, we utilize two primary metrics: Cosine Similarity (Cos) and F0 Frame Error (FFE).
Cosine Similarity is employed to evaluate the resemblance in singer identity between the synthesized singing voice and the audio prompt. This is achieved by calculating the average cosine similarity between the embeddings of the synthesized singing voices and the GT singing voices, offering an objective measure of the singer similarity. Specifically, we extract singer embeddings using the WavLM model \cite{chen2022wavlm}, which has been fine-tuned for speaker verification \footnote{https://huggingface.co/microsoft/wavlm-base-plus-sv}.
In addition, we use F0 Frame Error (FFE), which combines two key aspects: voicing decision errors and F0 errors. FFE serves as a comprehensive metric, effectively capturing crucial information related to the synthesis quality.

\section{Details of Baselines}
\label{sec: appendix4}

\textbf{StyleTTS 2} \citep{li2024styletts} integrates style diffusion and adversarial training with large speech language models (SLMs) for high-quality text-to-speech synthesis. It represents style as a latent variable using diffusion models. We use and revise their official code \footnote{https://github.com/yl4579/StyleTTS2}.

\textbf{CosyVoice} \citep{du2024cosyvoice} encodes speech with supervised semantic tokens derived from a multilingual speech recognition model and employs vector quantization in the encoder. It uses a large language model (LLM) for text-to-token generation and a conditional flow matching model for speech synthesis. We use and revise their official code \footnote{https://github.com/FunAudioLLM/CosyVoice}.

\textbf{VISinger 2} \citep{zhang2022visinger} combines digital signal processing (DSP) with VISinger to improve synthesis quality. By incorporating a DSP synthesizer with harmonic and noise components, it generates both periodic and aperiodic signals from the latent representation. The modified HiFi-GAN produces high-quality singing voices. We use their official code \footnote{https://github.com/zhangyongmao/VISinger2}.

\textbf{TCSinger} \citep{zhang2024tcsinger} introduces three key components: 1) a clustering style encoder to condense style information, 2) a Style and Duration Language Model (S\&D-LM) to predict style and phoneme duration, and 3) a style-adaptive decoder for enhanced detail in the singing voice. We use their official code \footnote{https://github.com/AaronZ345/TCSinger}.

\section{Details of Results}
\label{sec: appendix5}

\subsection{CFG}

Following previous CFG works\citep{jiang2025megatts}, we experiment with various parameter settings to verify the $\gamma$ value in the CFG, as shown in Table \ref{tab: cfg}. 
For style transfer and style control evaluation, we conduct CMOS assessments.
When $\gamma = 1$, $v_{cfg}$ becomes equivalent to the original formulation $v_t(x, t|C,P;\theta)$. 
When $\gamma$ ranges from 1 to 3, the generated singing voices are more consistent with the styles of the audio or textual prompts.
However, when $\gamma$ exceeds 5, the styles become exaggerated and unnatural, introducing artifacts and degrading the overall audio quality. This negatively impacts CMOS-Q.
By setting $\gamma = 3$, we achieve improved generation quality and ensure better style control.

\begin{table}[t]
\centering
\small
\begin{tabular}{lcccc}
\toprule
\multirow{2}{*}{\bfseries{$\gamma$}} & \multicolumn{2}{c}{\bfseries{Style Transfer}} & \multicolumn{2}{c}{\bfseries{Style Control}}\\ \cmidrule(lr){2-3} \cmidrule(lr){4-5}
 & CMOS-Q & CMOS-S & CMOS-Q & CMOS-C \\
\midrule
1 & -0.26 & -0.22 & -0.25 & -0.31 \\
2  & -0.21  & -0.14 & -0.19 & -0.25 \\
3 & 0.00  & 0.00 & 0.00 & 0.00 \\
5 & -0.25  & -0.02 & -0.27 & -0.02 \\
\bottomrule   
\end{tabular}
\caption{
Ablation study on CFG.
}
\label{tab: cfg}
\end{table}

\subsection{Cus-MOE}

\begin{table}[t]
\vskip 0.15in
\begin{center}
\begin{small}
\scalebox{1}{
\begin{tabular}{lcc}
\toprule
$ \bf Expert $ & \bf CMOS-Q \\
\midrule
1 & -0.53 \\
2 & -0.41 \\
3 & -0.20 \\
4 & \bf 0.00 \\
5 & 0.03 \\
\bottomrule
\end{tabular}}
\caption{Ablation study for Cus-MOE.}
\label{tab: moe}
\end{small}
\end{center}
\vskip -0.1in
\end{table}

Following previous audio generation works with MOE \citep{zhang2025versatile,zhang2025isdrama}, to examine the impact of the expert count within the Cus-MOE architecture, we conduct a series of style control experiments, varying the number of experts and assessing the results. These findings are summarized in Table \ref{tab: moe}.
We employ CMOS evaluation to quantify perceptual differences in the generated singing voices. Our analysis indicates a trend where the quality of generation improves with an increase in the number of experts. Specifically, increasing the expert count from the baseline configuration leads to noticeable improvements. However, this improvement plateaus after four experts.
The diminishing returns observed beyond this point can be attributed to several factors:
1) Model complexity: A larger number of experts may introduce redundant parameters, increasing model complexity and potentially hindering effective training convergence, thus making the learning process less efficient.
2) Computational overhead: Employing more experts significantly raises computational demands during both training and inference. However, the performance benefits do not scale proportionally with this increased resource consumption.
Considering the trade-off between performance and computational efficiency, we opt for a configuration of four experts per group. This choice strikes a balance between synthesis quality and resource utilization.

\end{document}